\documentclass{article}
\usepackage{ijcai19}
\usepackage[T1]{fontenc} 
\usepackage{times}
\usepackage{soul}
\usepackage{url}
\usepackage[hidelinks]{hyperref}
\usepackage[utf8]{inputenc}
\usepackage[small]{caption}

\usepackage{graphicx}
\usepackage{subcaption}
\usepackage{amsmath}
\usepackage{amssymb}
\usepackage{booktabs}
\DeclareMathOperator*{\E}{\mathbb{E}}

\title{CLVSA: A Convolutional LSTM Based Variational Sequence-to-Sequence Model with Attention for Predicting Trends of Financial Markets}

\author{
Jia Wang$^1$\and
Tong Sun$^1$\and
Benyuan Liu$^1$\and
Yu Cao$^1$\And
Hongwei Zhu$^2$\\
\affiliations
$^1$Department of Computer Science, University of Massachusetts Lowell \\
$^2$Department of Operations and Information Systems, University of Massachusetts Lowell \\
\emails
\{jwang, tsun, bliu, ycao\}@cs.uml.edu,
harry\_zhu@uml.edu
}

\begin{document}

\maketitle

\begin{abstract}
Financial markets are a complex dynamical system. The complexity comes from the interaction between a market and its participants, in other words, the integrated outcome of activities of the entire participants determines the markets trend, while the markets trend affects activities of participants. These interwoven interactions make financial markets keep evolving. Inspired by stochastic recurrent models that successfully capture variability observed in natural sequential data such as speech and video, we propose CLVSA, a hybrid model that consists of stochastic recurrent networks, the sequence-to-sequence architecture, the self- and inter-attention mechanism, and convolutional LSTM units to capture variationally underlying features in raw financial trading data. Our model outperforms basic models, such as convolutional neural network, vanilla LSTM network, and sequence-to-sequence model with attention, based on backtesting results of six futures from January 2010 to December 2017. Our experimental results show that, by introducing an approximate posterior, CLVSA takes advantage of an extra regularizer based on the Kullback-Leibler divergence to prevent itself from overfitting traps.
\end{abstract}

\section{Introduction}\label{Introduction}


\begin{figure}
  \begin{subfigure}[b]{0.21\textwidth}
    \includegraphics[width=\textwidth]{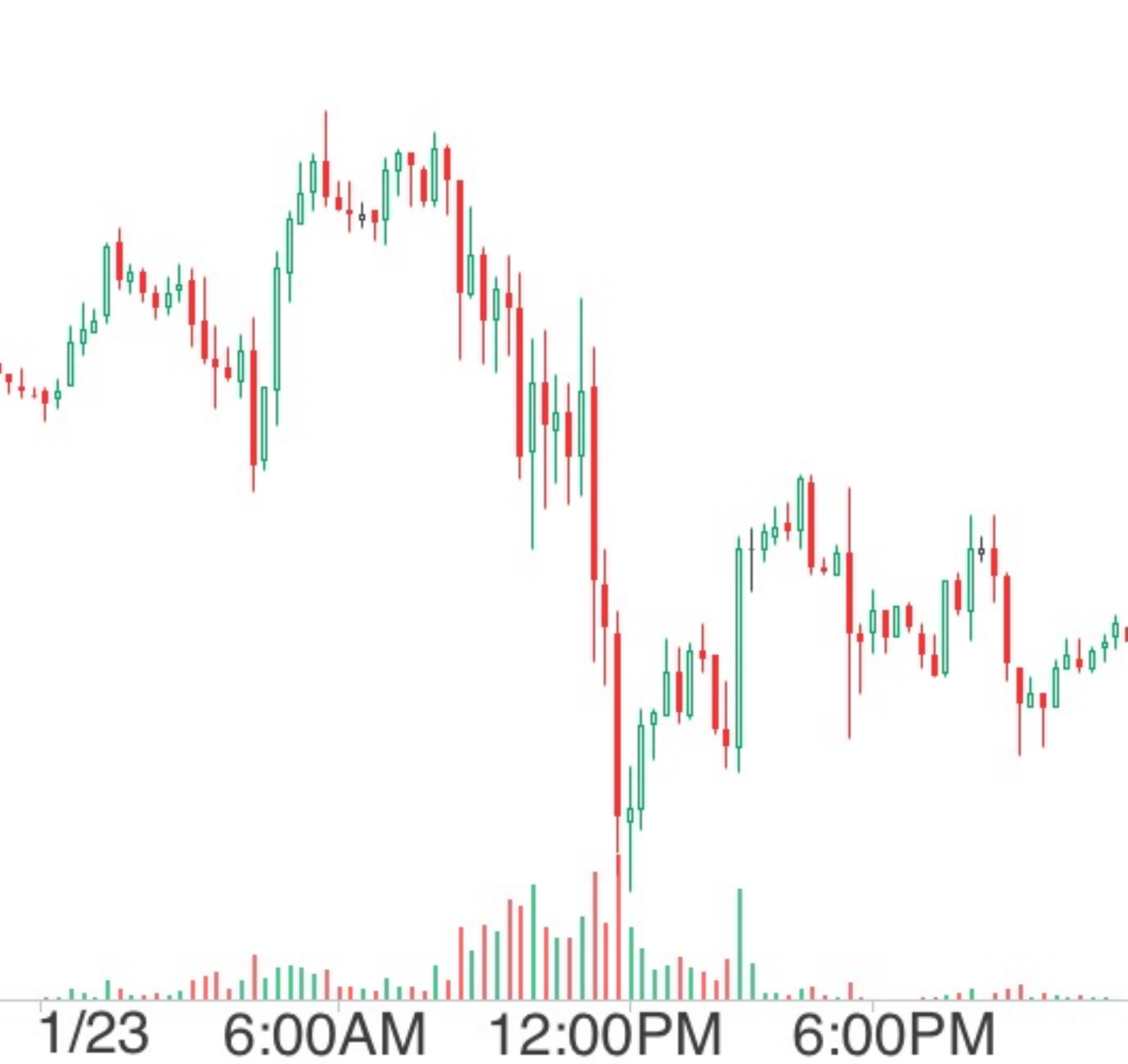}
    \caption{January-01-23, 2019}
  \end{subfigure}
  \quad
  \begin{subfigure}[b]{0.21\textwidth}
    \includegraphics[width=\textwidth]{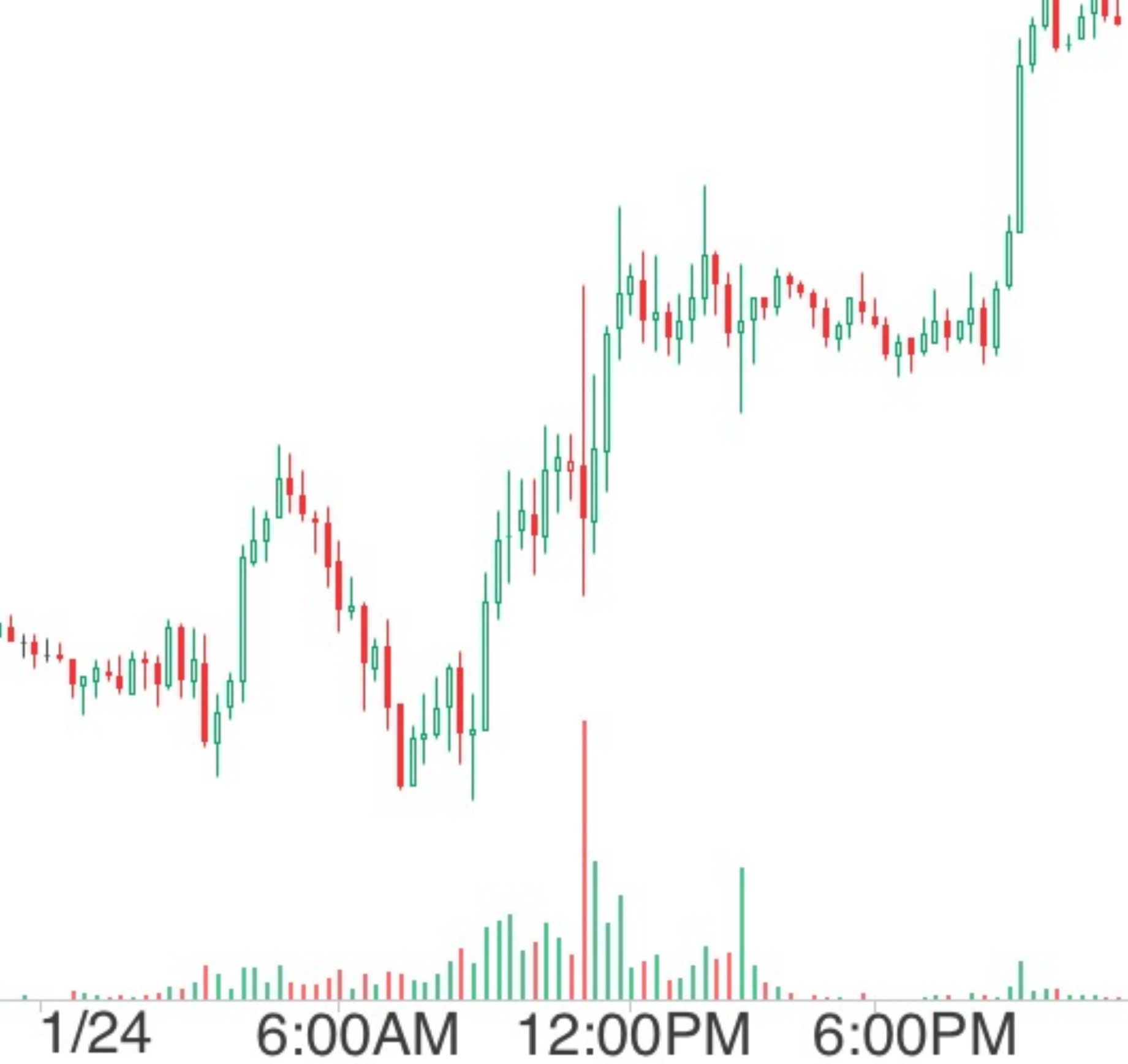}
    \caption{January-01-24, 2019}
  \end{subfigure}
  \caption{Two daily charts of Light Sweet Crude Oil (CL) futures. The daily pattern of trading volume (the bottom bar chart) has a visible regularity (see Section \ref{section:seq2seq} for more details).}\label{fig:yahoo}
\end{figure}

Predicting trends of financial markets is a very challenging task. Similar to other natural sequential data such as those of speech, text, and video, financial trading data contains latent temporal features that may reflect underlying market trends and financial patterns. Traditional methods for financial markets predictions, such as technical analysis/technical indicators \cite{achelis2001technical}, utilize charting and mathematical tools to seek latently profitable patterns from trading data. However, the difference between financial markets and other natural sequential events is that, the evolution of financial markets is mainly caused by the collective behavior of market participants rather than governed by natural physical laws. The adaptive market hypothesis proposed by \cite{lo2004adaptive} attributes the evolution of financial markets to the following reasons: i) Market participants act in their self-interest. ii) Market participants make mistakes due to loss aversion, overconfidence, and overreaction, etc, but they learn and adapt from mistakes. iii) The game between market participants drives markets to keep adapt and evolve over time. The above arguments provide a plausible explanation for why informative features for financial markets predictions are often transitory and difficult to extract.

How to capture latent features from trading data is the key to build robust predicting systems for financial markets. Some research, such as \cite{kim2003financial,fernandez2003technical,dixon2016classification}, use machine learning algorithms (e.g. SVM, Nearest Neighborhood, and Feed-forward networks) to capture latent features from technical indicators. Although technical indicators have been widely used by market participants, these methods may inevitably introduce human biases into models. Another popular sources for extracting latent features are market-related texts and information, such as reports, news, and tweets. \cite{ding2015deep,sun2017predicting,xu2018stock} use natural language processing approaches to predict price movement of stock markets, however, they are not applicable in high-frequency trading systems due to the time-lag property of these sources, and would be biased by fake information. 

In this paper, we propose a hybrid model, named Convolutional LSTM based Variational Sequence-to-Sequence model with Attention (CLVSA), to dynamically extract latent representations of trends of financial markets directly from trading data. We use convolutional LSTM units \cite{xingjian2015convolutional} to capture the characteristics of financial trading data, specifically, local features by convolutions and temporal features by recurrent LSTM networks. We split trading data into individual days (see Section \ref{ssec:datasets} for more details), use the encoder and decoder of the sequence-to-sequence (Seq2Seq) framework \cite{sutskever2014sequence} to handle the data segments of two consecutive days. We introduce Seq2Seq for the following reasons: i) Trading volume is an important measure for the relative worth of markets movement, and it has a daily regularity as shown in Figure \ref{fig:yahoo} that the quantity of trades tend to be high from 9am to 2pm (see Section \ref{section:seq2seq} for more details). Seq2Seq takes advantage of the above daily regularity of trading volume to capture temporal features.  ii) The decoder may obtain extra temporal features from the encoder. We also apply self-attention \cite{cheng2016long} and inter-attention mechanism \cite{bahdanau2014neural} to assist Seq2Seq in spotting latent temporal features in trading data.

Since labels are only samples of the market movements, they do not capture the fine-grained evolvement of market dynamics. As a result, the objective function based on labels can not completely reflect the adaptability of financial markets, and thus may cause overfitting. To address this issue, we introduce a backward decoder as the approximate posterior, to generate Kullback-Leibler divergence (KLD) as an extra regularizer for the model optimization.



We summarize our contributions as follows:
\begin{enumerate}
\item To our best knowledge, this work is the first attempt using a Seq2Seq framework with convolutional LSTM units and attention mechanisms to predict trends of financial market movement. The model takes only raw financial trading data as input without any intermediate human interventions so that it is a pure end-to-end approach.
\item We introduce an extra K-L divergence based regularizer to address the overfitting issue caused by the fact that labels are not able to perfectly represent the intrinsic characteristics of financial trading data. We compare the variational and non-variational versions of our model to demonstrate that the extra regularizer can significantly improve the robustness of our prediction systems.
\item We train our model with 8-year-long trading data of six futures and evaluate it with both financial and machine learning criteria. Our experimental results show that CLVSA provides the highest and most robust returns for all the six futures, compared to the basic models, such as deep convolutional neural network, and vanilla LSTM networks and Seq2Seq framework with attention.
\end{enumerate}

The remainder of the paper is organized as follows. Related work on financial market prediction with machine learning methods and related work on sequential learning methods is presented in Section \ref{related work}. The necessary background and the architecture of CLVSA are presented in Section \ref{model}. The datasets, experimental setup, and the criteria of both finance and machine learning that we use in this research are described in Section \ref{Experimental setup}. The experiments results and discussion are presented in Section \ref{Experimental results}, followed by concluding remarks in Section \ref{Conclusion}.

\section{Related Work}\label{related work}

\begin{figure}
\centering
\includegraphics[width=0.95\linewidth]{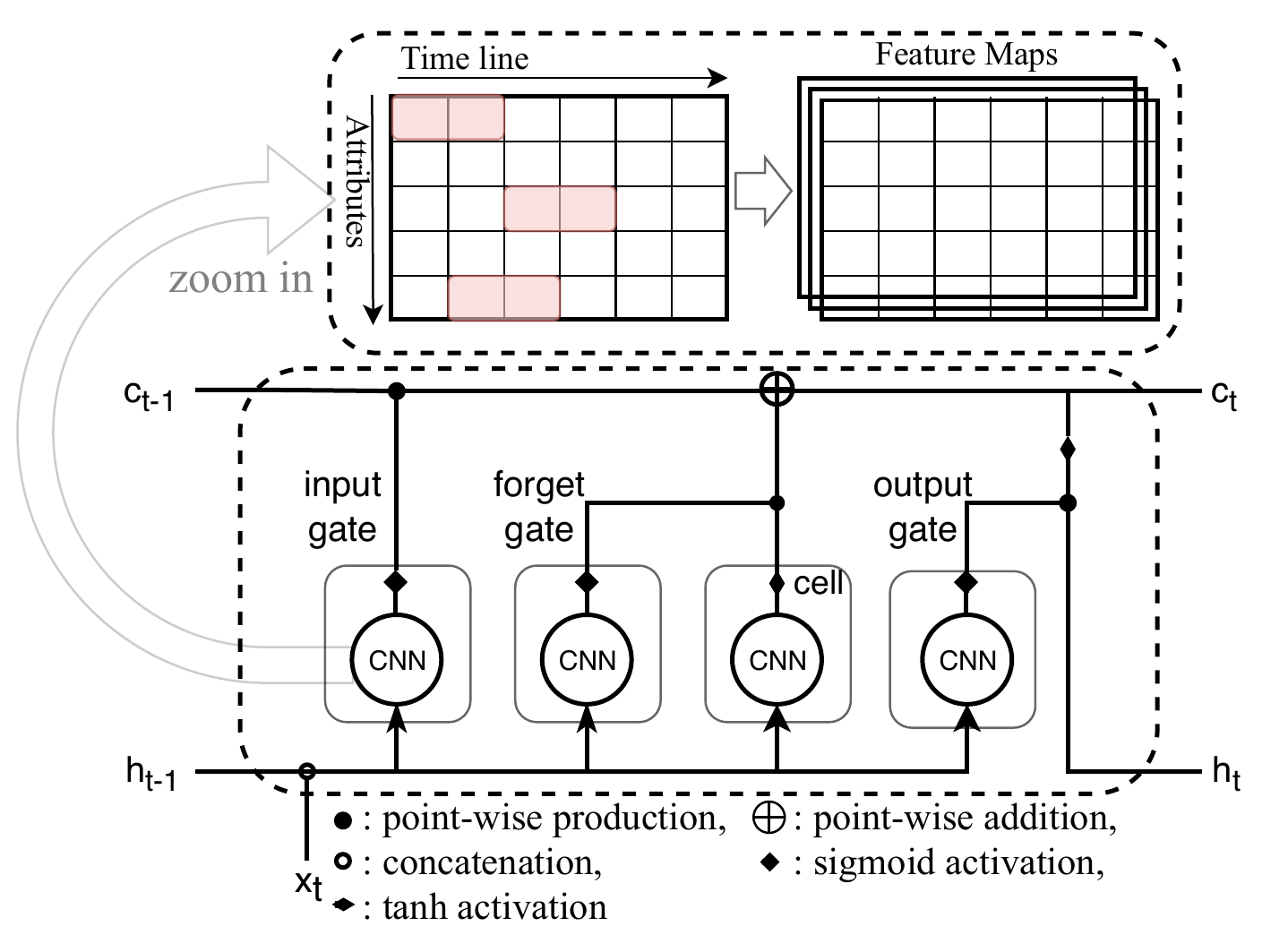}
\caption{The architecture of the Convolutional LSTM unit}
\label{fig:convLSTM}
\end{figure}

Although traditional predicting approaches such as technical analysis/indicators have existed for over hundreds of years, automated trading systems based on pattern recognition and machine learning have been popular since the 1990s. Various algorithms, such as SVM, nearest-neighbour, decision trees, and feed-forward neural networks have been applied to predict stocks, foreign exchange, and commodity futures markets \cite{kim2003financial,fernandez2003technical,rechenthin2014machine,dixon2016classification}. All the aforementioned work use technical indicators as input features. Since 2010s, more research utilizes the power of deep learning algorithms to predict financial markets instead of technical indicators. \cite{ding2015deep,wang2018financial} use deep convolutional neural networks to capture potential trading features from financial events and financial trading data, respectively.
\cite{zhang2017stock} proposes a variant of LSTM enhanced by discrete fourier transform to discover Multi-Frequency Trading Patterns. \cite{xu2018stock} use a stochastic recurrent model (SRM) with an extra discriminator and attention mechanisms to address the adaptability of stock markets, it is, however, driven by language-based data (tweets) rather than financial trading data. \cite{bacoyannis2018idiosyncrasies} proposes an approach based on reinforcement learning to model automated data-centric decision makers in quantitative finance.

Bonding the local features extraction ability of deep convolutional neural networks with the temporal features retention of LSTM, convolutional LSTM proposed by \cite{xingjian2015convolutional} has been applied in many fields such as weather forecasting \cite{xingjian2015convolutional}, image compression \cite{toderici2015variable}, and general algorithmic tasks (e.g. binary addition) \cite{kaiser2015neural}. The sequence-to-sequence framework proposed by \cite{sutskever2014sequence} achieves a significantly success in neural machine translation tasks, it is enhanced subsequently by inter-attention \cite{bahdanau2014neural} and self-attention \cite{cheng2016long}. \cite{kingma2013auto,rezende2014stochastic} proposes variational auto-encoder (VAE) that uses the encoder to form the approximate posterior, then trains the generative decoder to approximate the inputs of the encoder with variational lower bound and KLD. SRM \cite{bayer2014learning,goyal2017z} extends the basic idea of VAE into recurrent networks, using backward recurrent neural networks as the approximate posterior instead. Although SRM is widely used in natural language processing and speech recognition tasks, to our best knowledge to know, our research is the first effort to apply this method to deal with financial trading data.

\begin{figure*}
\centering
\includegraphics[width=.75\linewidth]{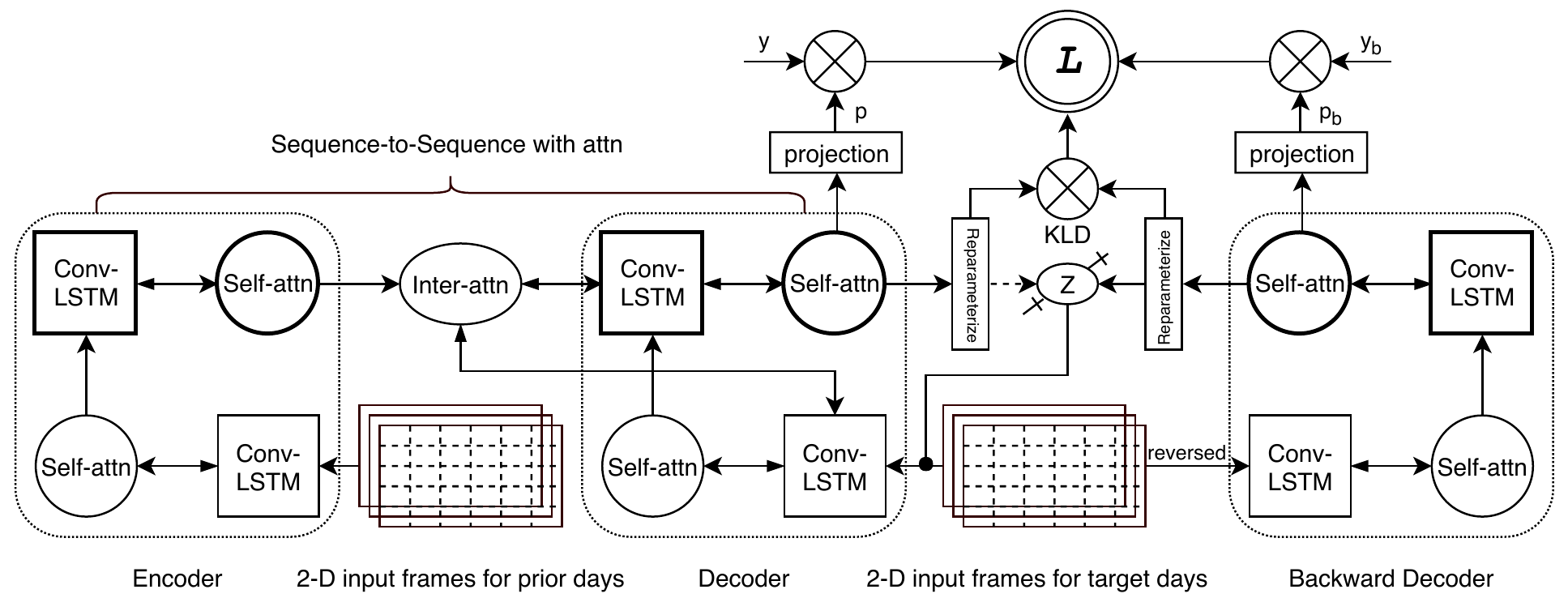}
\caption[small]{The architecture of CLVSA. 
The inter-attention module communicates the final layer of the encoder, to the both layers of the forward decoder. Since there exist interactions between convolutional LSTM units and attention modules, we use double-side arrows to connect them. Latent variable $z$ is given by the posterior (formed by the backward decoder) in the training procedure, and by the prior (formed by the decoder) in the evaluation/test procedure. We illustrate this process by a dashed- and solid-line arrow, and a selector symbol.}
\label{fig:CLVSA}
\end{figure*}
\section{Model Design}\label{model}

The architecture of our proposed model, CLVSA, is illustrated in Figure \ref{fig:CLVSA}. The 2-layer encoder and decoder of the sequence-to-sequence (Seq2Seq) framework take 2-D data frames of two consecutive days, respectively. Convolutional LSTM units (illustrated in Figure \ref{fig:convLSTM}) process 
2-D data frames by two steps: i) Convolutional kernels capture local features, ii) Based on the local features, LSTM networks capture temporal features with gated recurrent networks. In each layer of the encoder and decoder, we have a self-attention module to highlight parts of the sequence of daily data frames. We also have an inter-attention module to highlight parts of the first one of two consecutive days as the context of the second day. The backward decoder takes the reversed 2-D data frames of the second day for generating the KLD based regularizer.



\subsection{Convolutional LSTM Units}
All hidden layers in vanilla LSTM network \cite{hochreiter1997long} are fully connected layers, it works well for tasks such as speech recognition and natural language processing because we can perfectly map speeches and texts into trainable vector space \cite{mikolov2013distributed}. However, it struggles with the other tasks of which purely numeric data can not be tokenized. Convolutional LSTM units addresses this issue by replacing fully connected layers with convolutional kernels. Formally,
\begin{align*}
&i_{t} = \sigma(W_{i}*[X_{t},H_{t-1}] + b_{i}) \\
&f_{t} = \sigma(W_{f}*[X_{t},H_{t-1}] + b_{f}) \\
&o_{t} = \sigma(W_{o}*[X_{t},H_{t-1}] + b_{o}) \\
&C_{t} = f_{t} \circ C_{t-1} + i_{t} \circ \tanh(W_{c}*[X_{t},H_{t-1}] + b_{c}) \\
&H_{t} = o_{t} \circ \tanh(C_{t}),
\end{align*}
where $X_{t}$, $H_{t}$, $C_{t}$ denote the input data, hidden state, and cell state at time step $t$, respectively; $i_{t}$, $f_{t}$, $o_{t}$ denote the output of the input, forget, and output gate at time step $t$; $W_{i}$, $W_{f}$, $W_{o}$ denote the convolutional kernels of the input, forget, and output gate; $b_{i}$, $b_{f}$, $b_{o}$ denote the bias of the input, forget, and output gate. The symbol $*$ denotes the convolution operator and $\circ$ denotes the point-wise product operation. Since parameters sharing is the key factor for generalization of deep learning models, with the convolutional operations between the kernels and data, parameters of the kernels are shared among input data so that convolution is more powerful to extract generalized features from data.

However, we can not directly use the original design of convolutional LSTM units due to the property of financial trading data. Although trading data can be organized into 2-D frames, we can not apply 2-D convolutions because the rows of the 2-D frames contains different types of features including Open, High, Low, Close price, and Trading Volume. Instead we use a modified version of 1-D convolution: kernels only move horizontally across columns, the time line of 2-D data frames, while they are also shared among different types of data to guarantee parameters sharing. Note that we follow the design in \cite{xingjian2015convolutional} that we do not compress the size of input frames during convolutions, that means if input frames $X_{t}$ are 6x5 and output channels are 32, then the final hidden state output $H_{t}$ should be 6x5x32 and they are flattened to a vector as the input of the subsequent classifier layer.

\subsection{Sequence-to-Sequence Framework With Attention Mechanisms}\label{section:seq2seq}
There are two main motivations to use Seq2Seq framework with attention mechanisms (illustrated in Figure \ref{fig:CLVSA}) as the foundation of our model. First, trading volume has a significant rhythmicity. Take futures markets for instance, in every regular trading day, the quantity of trades keeps low in the midnight, starts to slowly increase in the early morning, reaches a plateau from 9am to 2pm around, and then quickly falls to the low ebb (as shown in Figure \ref{fig:yahoo}). Second, although longer sequences may provide more context information, training long sequences may make cross entropies too large for the model to converge due to the noisy and dynamic characteristics of financial markets. We thus split trading data into segments by individual days to fit the above regularity, then feed trading data of two consecutive days to the encoder and decoder, respectively.

We use the inter-attention mechanism \cite{bahdanau2014neural} to strengthen the connection between the encoder and decoder. The basic idea is that, at time step $t_{d}$ of the decoder, we concatenate its hidden state $hd_{t_{d}}$ with the weighted average of hidden states of the encoder. The weight of each time step $t_{e}$ of the encoder is determined by the similarity between $he_{t_{e}}$ and $hd_{t_{d}}$. Formally,
\begin{align*}
&ai_{t_{d}} = softmax(hd_{t_{d}}^TH_{e}), \;\; ci_{t_{d}} = \sum_{t_{e}}{ai_{t_{d}} he_{t_{e}}} \\
&hd^{'}_{t_{d}} = \tanh(Wi_{a} [hd_{t_{d}},ci_{t_{d}}]),
\end{align*}
where $ai_{t_{d}}$ is a vector containing weights for all time steps of the encoder with respect to time step $t_{d}$ of the decoder, $H_{e}$ is the whole hidden states of the encoder, $ci_{t_{d}}$ is the weighted average of all hidden states of the encoder with respect to $hd_{t_{d}}$, $h^{'}_{d_{t}}$ is the desired hidden state of the decoder at time step $t$ that contains the correlated information from the encoder. 

Trading volume is an important measure for the relative worth of market movement, time steps with high trading volume might carry more informative latent features. It is thus worth using self-attention mechanism \cite{cheng2016long} to highlight hot spots in both the encoder and decoder. Formally,
\begin{align*}
&as_{t} = softmax(h_{t}^TH^{'}_{1:t-1}), \;\; cs_{t} = \sum^{t-1}_{i=1}{as_{t_{i}} h_{t}} \\
&h^{'}_{t} = \tanh(Ws_{a} [h_{t},cs_{t}]),
\end{align*}
where $as_{t}$ is a vector containing weights for the hidden states of all previous time steps with respect to the hidden state at time step $t$, $cs_{t}$ is the weighted average of all hidden states of the previous time steps with respect to the hidden state at time step $t$, $h^{'}_{t}$ is the desired hidden state at  time step $t$, and $H^{'}_{1:t-1}$ represents $[h^{'}_{1}, ..., h^{'}_{t-1}]$. The content-base measurement method is dot production, which is same as the inter-attention mechanism.

\begin{table*}[!htp]
\centering
        \scalebox{0.72}{
            \begin{tabular}{*{19}{c|}}
                \cline{2-19}
                 & \multicolumn{3}{|c|}{CL}
                 & \multicolumn{3}{|c|}{NG}
                 & \multicolumn{3}{|c|}{GC}
                 & \multicolumn{3}{|c|}{S}
                 & \multicolumn{3}{|c|}{NQ}
                 & \multicolumn{3}{|c|}{ES}\\
                \cline{2-19}
                 & MAP & AAP & SP & MAP & AAP & SP & MAP & AAP & SP & MAP & AAP & SP & MAP & AAP & SP & MAP & AAP & SP \\
                \hline
                \multicolumn{1}{|c|}{$CNN_{s}$} & $43.8\%$ & $47.7\%$ & $1.55$ & $39.1\%$ & $37.0\%$ & $1.05$ & $35.6\%$ & $15.4\%$ & $0.67$ & $42.6\%$ & $44.3\%$ & $1.20$ & $45.1\%$ & $36.2\%$ & $1.54$ & $42.5\%$ & $33.8\%$ & $0.75$\\
                \multicolumn{1}{|c|}{$LSTM_{s}$} & $39.0\%$ & $41.4\%$ & $1.12$ & $40.0\%$ & $40.4\%$ & $1.23$ & $38.1\%$ & $21.1\%$ & $0.82$ & $40.1\%$ & $34.9\%$ & $1.13$ & $43.5\%$ & $32.2\%$ & $1.26$ & $40.0\%$ & $31.1\%$ & $0.62$\\
                
                \multicolumn{1}{|c|}{$Seq2Seq_{s}$} & $39.5\%$ & $39.4\%$ & $1.02$ & $39.6\%$ & $40.9\%$ & $1.11$ & $38.1\%$ & $20.8\%$ & $0.79$ & $40.2\%$ & $35.2\%$ & $1.17$ & $43.3\%$ & $32.7\%$ & $1.28$ & $40.3\%$ & $32.6\%$ & $0.67$ \\
                \hline
                \multicolumn{1}{|c|}{$CLSA$} & $47.8\%$ & $88.9\%$ & $2.69$ & $48.3\%$ & $60.1\%$ & $2.23$ & $43.7\%$ & $42.9\%$ & $1.08$ & $45.5\%$ & $54.2\%$ & $1.66$ & \pmb{$46.9\%$} & $41.8\%$ & $1.85$ & $44.3\%$ & $42.0\%$ & $1.02$ \\
                \hline
                \hline
                \multicolumn{1}{|c|}{$CLVSA$} & \pmb{$49.7\%$}  & \pmb{$113.0\%$} & \pmb{$3.99$} & \pmb{$50.7\%$} & \pmb{$73.7\%$} & \pmb{$3.01$} & \pmb{$45.5\%$} & \pmb{$50.2\%$} & \pmb{$1.58$} & \pmb{$48.0\%$} & \pmb{$57.6\%$} & \pmb{$1.81$} & $46.7\%$ & \pmb{$42.5\%$} & \pmb{$1.87$} & \pmb{$45.1\%$} & \pmb{$42.3\%$} & \pmb{$1.22$} \\
                \hline
                \end{tabular}
        }  
        \caption{Experimental results. MAP, AAP, and SP denote the mean average precision, average annual return, and Sharpe ratio, respectively. We ran each case five times to measure the robustness of each model. All values are average results of the repeated experiments.}
\label{tb:result}
\end{table*}

\subsection{Approximate Posterior and Stochasticity Injection}
As mentioned in Section \ref{Introduction}, the adaptability is the main challenge to model financial markets trends. We use labelling methods to classify trends of financial markets to the following three classes: Up, Flat, and Down. However, labelling methods inevitably miss some useful information, such as specific values of Open, High, Low, Close price and Volume at each time step. For example, one of the common labelling methods is to first compute logarithm return between the close price at time step $t$ and $t+1$, and then compare the logarithm return with a given threshold to define the label at time step $t$. The sum of cross entropies between ground truth and predictions of target sequential data thus only represents how different the predicted disperse series of close price movement deviate from the true one. It turns out that, the variability of latent features is ignored by the training process of supervised sequential models in that the variability is an abstract representation of the original data rather than the labels. In other words, the penalties from cross entropies between labels and softmax outputs may have biases due to the fact that labels can not perfectly represent the intrinsic adaptability of financial trading data. 

To alleviate the above biases, we propose an unsupervised method to generate an extra regularizer. Inspired by the variational auto-encoder \cite{kingma2013auto,rezende2014stochastic}, 
we introduce stochasticity into hidden states, the stochasticity can be trained by the reparameterization trick \cite{kingma2013auto} to form the prior distribution of latent features. Meanwhile, following recent work about stochastic recurrent models \cite{bayer2014learning,goyal2017z}, we introduce a backward decoder to form the approximate posterior, then we apply the K-L divergence between the prior and posterior distribution as the extra regularizer to the objective function. Formally,
\begin{equation}
\begin{gathered}\label{eq:alt_elbo}
\mathcal{L}(x, \theta, \phi)= \E_{p_{\phi}}(\log p_{\theta}(y|x,z)) - \\ D_{KL}(p_{\phi}(z|x,y)||p_{\theta}(z|x)),
\end{gathered}
\end{equation}
where $\theta$ and $\phi$ are parameters of random process for the prior and posterior, we use two neural networks to form them. The first term samples the approximate posterior $p_{\phi}(z|x,y)$ to attain the latent variable $z$, then use it as one of the conditions for the log likelihood $\log p_{\theta}(y|x,z)$. The approximate posterior is only used in the training procedure, $z$ is from the $p_{\theta}(z|x)$ in the evaluation and test procedure (shown as a dashed-line arrow to the selector for latent variable $z$ in Figure \ref{fig:CLVSA}). There would be no overfitting risks once KLD converges.

\subsection{The Objective Function}
The objective function of our model has the following three main parts: i) The cross entropies of predictions of the decoder, ii) The cross entropies of predictions of the backward decoder, iii) The K-L divergence between the prior and posterior distribution, which are formed by the decoder and the backward decoder, respectively.

To extend Equation (\ref{eq:alt_elbo}) into the sequential scenario, we have the objective function as follows,
\begin{equation}\nonumber
\begin{gathered}\label{eq:objective}
\mathcal{L}(x, \theta, \phi, \xi)= \sum_{t} \E_{p_{\phi}}(\log p_{\theta}(y_{t}|x_{1:t},z_{1:t}))
+ \alpha \log p_{\xi}(y^{'}_{t}|b_{t}) \\
- \beta D_{KL}(p_{\phi}(z_{t}|x_{T:t},y^{'}_{t})||p_{\theta}(z_{t}|x_{1:t},z_{1:t-1})) + \gamma L_{2},
\end{gathered}
\end{equation}
where $\xi$ denotes the parameters of the projection layers for the backward decoder, $b_{t}$ denotes the hidden state of the backward decoder at the $t_{th}$ time stamp. $x_{T:t}$ and $y^{'}_{t}$ are the reversed trading data and the corresponding labels for the backward decoder, respectively. $\alpha$ and $\gamma$ denotes the weight of $p_{\xi}(y^{'}_{t}|b_{t)}$ and $L_{2}$ regularizer, respectively. $\beta$ denotes the KLD annealing weight \cite{bowman2015generating}.
\section{Experimental Setup}\label{Experimental setup}

\subsection{Datasets}\label{ssec:datasets} We use historical trading data of four commodity futures and two equity index futures as the datasets in our research, including WTI Light Sweet Crude Oil (CL), Gold (GC), Natural Gas (NG), Soybeans (S), E-mini S\&P 500 (ES), and E-mini Nasdaq 100 (NQ). We collect these datasets for time period from January 2010 to December 2017 from online brokers such as Interactive Brokers. Each dataset consists of 334,000-404,000 5-minute trading records. Each record contains the following seven attributes: date, time, open price, high price, low price, close price, and trading volume. The trading volume represents the aggregation of all trades, the four prices represent the open price, the highest price, the lowest price, the close price at the corresponding 5-minute time interval, respectively. Note that the backward RNN to form the approximate posterior uses the reversed version of trading data.

We strictly follow the sequential order to split train/validation/test sets to avoid the data leaking problem \cite{rechenthin2014machine}, specifically, we use 3-year-long trading data as the train set, and the consecutive two-week data as the validation and test set, respectively. The train/validation/test set will shift forward by one week for the next training session. For the sake of convolutions, we incorporate six consecutive 5-minute trading records into a 30-minute 2-D data frame, the rows of which contain the following attributes: Open, High, Low, Close price, and Trading volume, the columns of which contain the six consecutive 5-minute time series. All data of the above five attributes have been normalized separately.

We use logarithmic return between two consecutive 30-min data frames as the labelling method, formally,
\begin{align*}
&b_{up} = \log((\mu_{c} + \lambda) / \mu_{c}), \;\; b_{down} = \log((\mu_{c} - \lambda) / \mu_{c}) \\
&y_{t} = 
\begin{cases}
1,   & \text{ if } \log{(p_{c_{t+1}}/p_{c_{t})}} > b_{up} \\
-1,  & \text{ if } \log{(p_{c_{t+1}}/p_{c_{t})}} < b_{down} \\
0,   & \text{ otherwise },
\end{cases}
\end{align*}
where $b_{up}$ and $b_{down}$ denote the threshold of Up and Down, respectively. $\lambda$ is a parameter for balancing the three classes to roughly 1:1:1. $p_{c_{t+1}}$ and $p_{c_{t}}$ denote the close price of the (t+1)-th and t-th 30-minute data frame, respectively.

\begin{figure*}[htp!]
\centering
\includegraphics[width=0.99\linewidth]{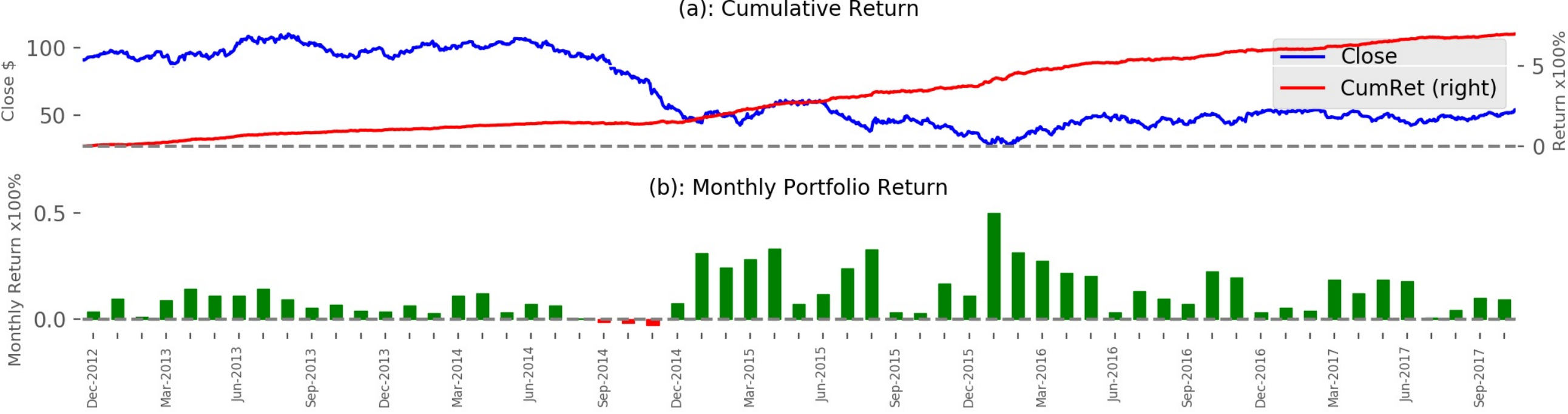}
\caption{The cumulative and monthly return of WTI Light Sweet Crude Oil (CL) futures by CLVSA.}
\label{fig:cr}
\end{figure*}

\subsection{Evaluation Criteria} We use mean average precision as the machine learning criterion for the 3-class classification task. 

In order to verify the potential predictive capability of our model for real-time market trading, we use a common but strict backtesting procedure as the financial criterion. The backtesting procedure has \$100,000 as the initial capital, enters/leaves markets according to the predictions, calculates the profit rate for each trading activity, and finally reports results of financial metrics, such as average annual return, and Sharpe ratio. The details of this procedure is that, The first Up or Down predicting label renders the test into the long trade or short trade, respectively. After that, once a turning point appears, the test leaves the current trade and then enters the next one. For example, the current trading status is the long trade but the next predicting label is Down, the trading strategy will leave the long trade and enter the short trade. We set high transaction costs to guarantee the strictness of our backtesting strategy, Take CL for instance, the bid-ask spread is \$0.01, the minimum price fluctuation is \$0.01, the multiplier is 1,000 times, and the commissions is \$2.75 at Interactive Brokers, the total transaction cost per contract in our backtest is up to \$85.5, which is two times higher than the average real-life transaction cost.

\subsection{Baselines and the Training Specification} Our real goal in this research is to test whether CLVSA achieves better performance than basic models, we choose the following approaches as the baselines: convolutional neural network ($CNN_{s}$), recurrent neural network with LSTM units ($LSTM_{s}$), sequence-to-sequence (Seq2Seq) model with attention mechanisms ($Seq2Seq_{s}$). We also use a convolutional LSTM based Seq2Seq model with attention mechanisms (CLSA) to test the beneficial effects of the KLD based regularizer. Note that, except for $CNN_{s}$, which follows the setting in \cite{wang2018financial},  all baseline approaches have the same structures as the corresponding components in CLVSA. Specifically, the number of the output channels of convolutional LSTM units is 32 and the size of convolution kernels is 3x1, both the encoder and the decoder of the Seq2Seq model are a stack of two convolutional LSTM units, and the discriminator consists of a 200-unit and 50-unit FC layer and a softmax layer. The backward RNN for approximate posterior in CLVSA has the same structure as the decoder. For the reparametrization tricks \cite{kingma2013auto}, The multi-variant Gaussian distributions for the prior and the posterior distribution are formed by a 512-unit and 256-unit FC layer, respectively.

We train all the models by Adam optimizer \cite{kingma2014adam} with a learning rate of 0.001. We use a TITAN RTX GPU with mini-batch size of 16 for all experiments. We set a dropout \cite{srivastava2014dropout} of 0.1 for all the FC layers that are activated by ReLU activation function, and set $\alpha$ and $\gamma$ by $2.5\mathrm{e}{-4}$ and $1\mathrm{e}{-5}$ for CLVSA. The KLD anealing weight linearly increases from 0 to 1 with iterations. We train all models for 1,000 iterations (17 epoches). For each iteration, we randomly sample mini batches from training sets.

\section{Experimental Results}\label{Experimental results}

We consider a total of 30 cases (five models and six datasets) in our experiments. We ran each case five times to test the robustness of the five models and the experimental results are shown in Table \ref{tb:result}. CLSVA achieves the best performance among the five models for all six futures. Compared to the three basic models, CLVSA outperforms them for mean average precision (MAP), average annual return (AAR), and Sharpe ratio (SR) by up to 11.6\%, 73.6\%, and 2.92, respectively. Compared to the non-variational version of our hybrid model CLSA, CLVSA outperforms it for mean average precision, average annual return, and Sharpe ration by 1.9\%, 24.1\%, and 1.30, respectively.

Our model achieves the best performance for the CL futures over the other futures due to the following two plausible reasons: i) CL is one of the most actively traded futures, its high volatility provides more opportunities for profitable tradings. ii) The labelling method we use is based on the logarithmic return of Close price so it is more appropriate for highly active futures such as CL. Figure \ref{fig:cr} illustrates experimental results for the CL futures with CLVSA model.
The cumulative return of the CL futures stays positive for all the months, and eventually achieves 600\%, and 56 out of the 59 months achieve positive monthly return, and no months have negative return lower than -5\%.

\begin{figure}
\centering
\includegraphics[width=0.9\linewidth]{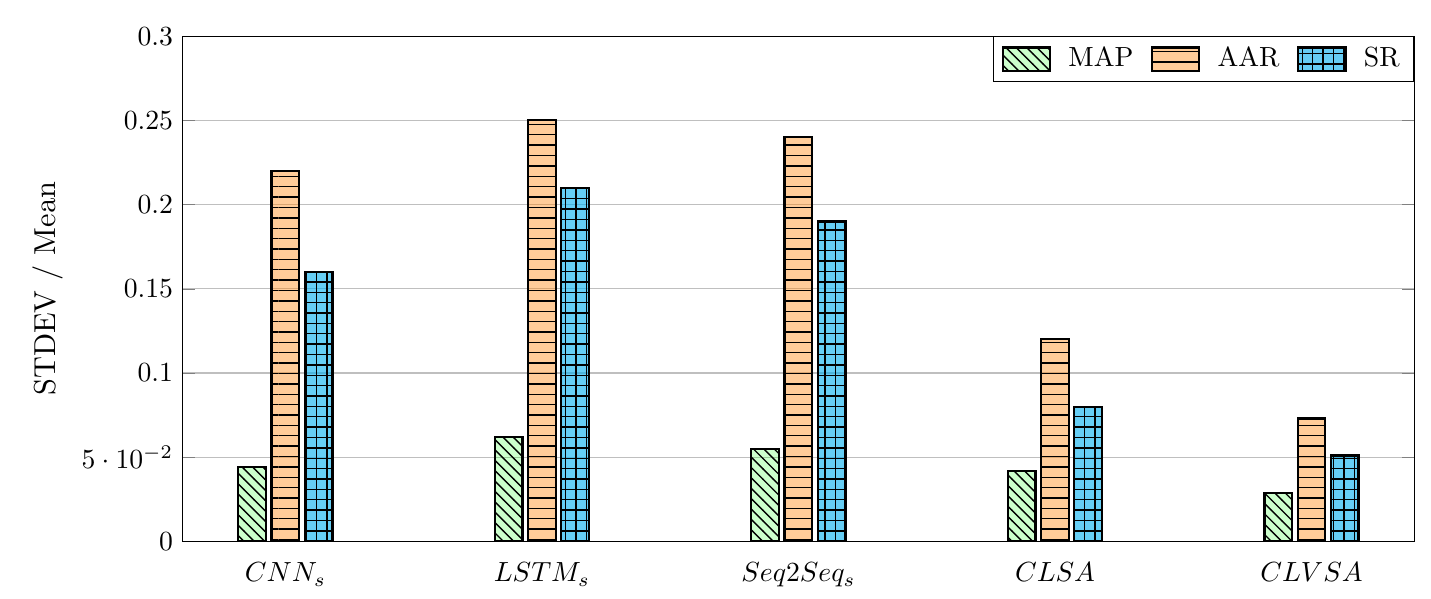}
\caption{The coefficient of variation for MAP, AAR, and SR of CL futures repeated experiments. Smaller values indicate stronger robustness. }
\label{fig:dm}
\end{figure}

We calculate the coefficient of variation for MAP, AAR, and SR over repeated experiments and the results for CL futures are shown in Figure \ref{fig:dm}. CLVSA achieves the smallest coefficient of variation for all the three criteria, which indicates that CLVSA provides more stable and robust predictions than other models. In particular, since CLVSA is the variational version of CLSA, which verifies that the extra regularizer efficiently prevents the overfitting problems and enable our model to capture more generalized temporal features. In contrast, the baseline models provide unstable results probably due to their incapability to capture latent features from raw trading data. Specifically, $CNN_{s}$ can capture local features in data frames, but it can not further attain temporal features from the context; $LSTM_{s}$ and $Seq2Seq_{s}$ struggle with the noisy characteristic of raw trading data although they perform well for sequential tasks.


We also observe the following phenomena that are worth discussing: i) Although higher MAP leads to higher AAR and SR, it dose not correlate well with these two financial criteria. This may be due to the fact that the true positives of class Up and Down are more important than the ones of class Flat for AAR and SR, but the three classes are equally important for MP. ii) All the models did a good job to predict class Flat, but the basic models struggle with distinguishing between class Up and Down. Expect for the limitations of the baseline models (e.g. $CNN_{s}$ can not capture temporal features, $LSTM_{s}$ and $Seq2Seq_{s}$ can not effectively capture local features), another plausible explanation is that the trading volume in our datasets is the aggregation of Up and down volume, which may cause models to confuse. 

\section{Conclusion}\label{Conclusion}

In this paper, we propose and develop a hybrid model named CLVSA to predict trends of financial markets. It consists of convolutional LSTM units, the sequence-to-sequence framework with self- and inter-attention mechanisms, and an extra backward decoder to address the adaptability of financial markets. We use 8-year-long trading data to train our model, 5-time repeated experiments for 30 cases verify that CLVSA can extract latent temporal features more effectively than the baseline models such as deep convolutional neural networks and the vanilla sequence-to-sequence model. Experimental results show that our model significantly outperforms the baselines, providing higher and more robust returns.

\end{document}